\documentclass[prc,preprintnumbers,twocolumn]{revtex4}
\usepackage{graphicx}
\usepackage{amssymb}
\usepackage{amsmath}

\begin{document}

\title{Sub-barrier capture reactions with $^{16,18}$O beams}
\author{V.V.Sargsyan$^1$, G.G.Adamian$^{1}$, N.V.Antonenko$^1$, W. Scheid$^2$, and  H.Q.Zhang$^3$
}
\affiliation{$^{1}$Joint Institute for Nuclear Research, 141980 Dubna, Russia\\
$^{2}$Institut f\"ur Theoretische Physik der
Justus--Liebig--Universit\"at,
D--35392 Giessen, Germany\\
$^{3}$China Institute of Atomic Energy, Post Office Box 275, Beijing 102413,  China
}
\date{\today}

\begin{abstract}
Various   sub-barrier capture reactions with beams $^{16,18}$O
are treated   within the quantum diffusion approach. The role of neutron transfer in these capture
reactions is discussed.

\end{abstract}


 \maketitle

%\section{Introduction}
%

The purpose of this Brief Report is the theoretical explanation of the sub-barrier capture reactions
$^{16,18}$O+$^{52,50}$Cr,
%$^{16,18}$O+$^{60,58}$Ni,
$^{16,18}$O+$^{76,74}$Ge,
$^{16,18}$O+$^{94,92}$Mo, and
$^{16,18}$O+$^{112,114,118,120,124,126}$Sn.
Within the quantum diffusion approach~\cite{EPJSub,EPJSub1}
we try to  answer  the question how strong
 the influence of neutron transfer  in these  capture reactions.
This study is important for future  experiments indicated in Ref.~\cite{Jia}.

In the quantum diffusion approach~\cite{EPJSub,EPJSub1}
the collisions of  nuclei are described with
a single relevant collective variable: the relative distance  between
the colliding nuclei. This approach takes into consideration the fluctuation and dissipation effects in
collisions of heavy ions which model the coupling with various channels
(for example, coupling of the relative motion with low-lying collective modes
such as dynamical quadrupole and octupole modes of the target and projectile~\cite{Ayik333}).
We have to mention that many quantum-mechanical and non-Markovian effects accompanying
the passage through the potential barrier are taken into consideration in our
formalism~\cite{EPJSub,EPJSub1,PRCPOP}.
The  nuclear deformation effects
are taken into account through the dependence of the nucleus-nucleus potential
on the deformations and mutual orientations of the colliding nuclei.
To calculate the nucleus-nucleus interaction potential $V(R)$,
we use the procedure presented in Refs.~\cite{EPJSub,EPJSub1}.
For the nuclear part of the nucleus-nucleus
potential, the double-folding formalism with
the Skyrme-type density-dependent effective
nucleon-nucleon interaction is used.
With this approach many heavy-ion capture
reactions at energies above and well below the Coulomb barrier have been
successfully described~\cite{EPJSub,EPJSub1,PRCPOP}.
One should stress that the diffusion models, which  include the quantum statistical effects,
were also treated in Refs.~\cite{Hofman}.

Following the hypothesis of Ref.~\cite{Broglia},
we assume that the sub-barrier capture
in the reactions under consideration
mainly  depends  on the two-neutron
transfer with the  positive  $Q_{2n}$-value.
Our assumption is that, just before the projectile is captured by the target-nucleus
(just before the crossing of the Coulomb barrier) which is a slow process,
the  $2n$-transfer ($Q_{2n}>0$) occurs   that can lead to the
population of the excited collective
states in the recipient nucleus~\cite{SSzilner}.
So, the motion to the
$N/Z$ equilibrium starts in the system before the capture because it is energetically favorable
in the dinuclear system in the vicinity of the Coulomb barrier.
For the reactions  considered,
the average change of mass asymmetry is related to the two-neutron
transfer. In  these reactions  the
$2n$-transfer channel is more favorable than $1n$-transfer channel ($Q_{2n}>Q_{1n}$).
Since after the $2n$-transfer the mass numbers,  the deformation parameters
of the interacting nuclei, and, correspondingly, the height $V_b=V(R_b)$
and shape of the Coulomb barrier are changed,
one can expect an enhancement or suppression of the capture.
If  after the neutron transfer the deformations of interacting nuclei increase (decrease),
the capture probability increases (decreases).
If  after the transfer the deformations of interacting nuclei do not change,
there is no effect of the neutron transfer on the capture.
This scenario was verified in the description of many reactions~\cite{EPJSub1}.
\begin{figure}[h]
\vspace*{-0.cm}
\centering
\includegraphics[angle=0, width=0.9\columnwidth]{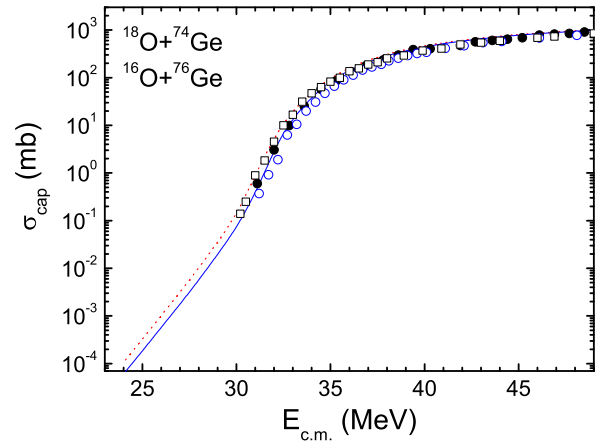}
\vspace*{-0.2cm}
\caption{The calculated (solid line) capture cross sections versus $E_{\rm c.m.}$ for the reactions
$^{16}$O+$^{76}$Ge  and $^{18}$O+$^{74}$Ge (the curves coincide).
For the $^{18}$O+$^{74}$Ge reaction,
the calculated capture cross sections without
the neutron transfer are shown by dotted line.
The experimental data  for the reactions $^{16}$O+$^{76}$Ge (open circles) and $^{18}$O+$^{74}$Ge (open squares)
are from Ref.~\protect\cite{Jia}.
The experimental data  for the $^{16}$O+$^{76}$Ge reaction (solid circles)
are from Ref.~\protect\cite{16OAGe}.
}
\label{1_fig}
\end{figure}
\begin{figure}[h]
\vspace*{0.4cm}
\centering
\includegraphics[angle=0, width=0.9\columnwidth]{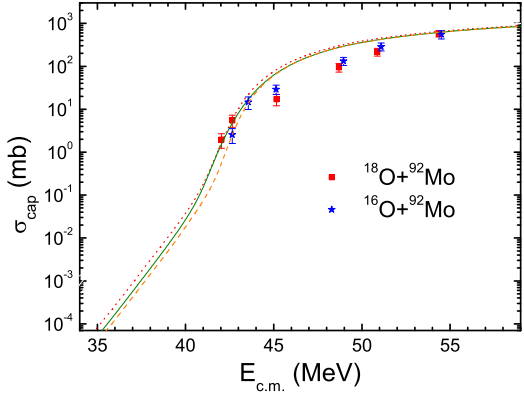}
\vspace*{-0.4cm}
\caption{The calculated  capture cross sections versus $E_{\rm c.m.}$ for the reactions
$^{16}$O+$^{92}$Mo (dashed line) and $^{18}$O+$^{92}$Mo (solid line).
For the $^{18}$O+$^{92}$Mo reaction,
the calculated capture cross sections without
the neutron transfer are shown by dotted line.
The experimental data  for the reactions
$^{16}$O+$^{92}$Mo (solid stars) and $^{18}$O+$^{92}$Mo (solid squares)
are from Ref.~\protect\cite{AO92Mo}.
}
\label{2_fig}
\end{figure}

All calculated results are obtained with the same set of parameters
 as in Ref.~\cite{EPJSub}
and are rather insensitive
to the reasonable variation of them~\cite{EPJSub,EPJSub1}.
Realistic friction coefficient in the momentum
$\hbar\lambda$=2 MeV
is used which
is close to those calculated within the mean field approaches~\cite{obzor}.
The parameters of the nucleus-nucleus interaction potential $V(R)$
are adjusted to describe the experimental
data at energies above the Coulomb barrier corresponding to spherical nuclei.
The absolute values of the quadrupole deformation parameters $\beta_2$
of even-even deformed nuclei are taken from Ref.~\cite{Ram}.
In Ref.~\cite{Ram} the quadrupole
deformation parameters $\beta_2$ are given for the first excited
2$^{+}$ states of nuclei. For the  nuclei deformed in the
ground state, the $\beta_2$ in 2$^{+}$ state is similar
to the $\beta_2$ in the ground state and we use $\beta_2$
from Ref.~\cite{Ram} in the calculations.
For the double magic   nucleus $^{16}$O,
in the ground state
we take $\beta_2=0$.
Since there are  uncertainties in the definition of the values of $\beta_2$
in  light- and medium-mass nuclei,
one can extract the
quadrupole deformation parameters of
these  nuclei from a comparison
of the calculated capture cross sections with the existing experimental data.
By describing the reactions
$^{18}$O+$^{208}$Pb,
where there are no neutron transfer channels with positive $Q$-values,
we extract $\beta_2=0.1$ for the ground-state
of  $^{18}$O~\cite{EPJSub1}.
This extracted value is used in our calculations.

\begin{figure}[h]
\vspace*{-0.cm}
\centering
\includegraphics[angle=0, width=0.9\columnwidth]{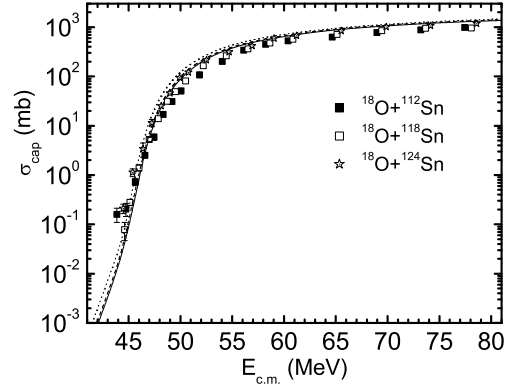}
\vspace*{-0.2cm}
\caption{The calculated  capture cross sections versus $E_{\rm c.m.}$ for the reactions
$^{16}$O+$^{114}$Sn  and $^{18}$O+$^{112}$Sn (solid line),
$^{16}$O+$^{120}$Sn and
$^{18}$O+$^{118}$Sn (dashed line), $^{16}$O+$^{126}$Sn
and $^{18}$O+$^{124}$Sn (dotted line).
The calculated results
for the reactions
$^{16}$O+$^{114,120,126}$Sn and $^{18}$O+$^{112,118,124}$Sn  coincide, respectively.
The experimental data  for the  reactions $^{18}$O+$^{112}$Sn (solid squares), $^{18}$O+$^{118}$Sn (open squares),
and $^{18}$O+$^{124}$Sn (open stars)
are from Ref.~\protect\cite{AOASn}.
}
\label{3_fig}
\end{figure}
\begin{figure}[h]
\vspace*{-0.cm}
\centering
\includegraphics[angle=0, width=0.9\columnwidth]{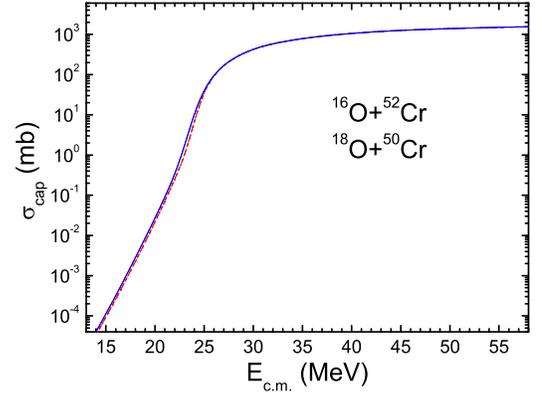}
\vspace*{-0.2cm}
\caption{The calculated  capture cross sections versus $E_{\rm c.m.}$ for the reactions
$^{16}$O+$^{52}$Cr (dashed line) and $^{18}$O+$^{50}$Cr (solid line).}
\label{4_fig}
\end{figure}

Figures  1-4 show the capture excitation function  for the reactions
$^{16,18}$O+$^{76,74}$Ge,
$^{16,18}$O+$^{94,92}$Mo,
$^{16,18}$O+$^{114,112,120,118,126,124}$Sn,  and $^{16,18}$O+$^{52,50}$Cr
as a function of  bombarding energy.
One can see a relatively good agreement between the
calculated results and the experimental data~\cite{Jia,16OAGe,AO92Mo,AOASn}
for the reactions
$^{16}$O+$^{76}$Ge,
$^{16,18}$O+$^{92}$Mo, and
$^{18}$O+$^{112,118,124}$Sn.
The $Q_{2n}$-values for the $2n$-transfer
processes are positive (negative) for all reactions with $^{18}$O ($^{16}$O).
Thus, the neutron transfer can be important for the reactions with  the $^{16}$O beam.
However, our results show that cross sections for reactions $^{16}$O+$^{76}$Ge ($^{16}$O+$^{114,120,126}$Sn,$^{52}$Cr)
and $^{18}$O+$^{74}$Ge ($^{18}$O+$^{114,118,124}$Sn,$^{50}$Cr) are very similar.
The reason of such behavior is that after the $2n$-transfer in the system $^{18}$O+$^{A-2}$X$\to ^{16}$O+$^{A}$X the deformations
remain to be  similar.
As a result, the corresponding Coulomb barriers of the  systems
$^{18}$O+$^{A-2}$X and $^{16}$O+$^{A}$X
 are almost the same and, correspondingly, their capture cross
sections  coincide.
The similar behaviour
was observed in the recent experiments $^{16,18}$O+$^{76,74}$Ge~\cite{Jia}.

One can see in Figs. 1-4 that at energies above   and near  the Coulomb barrier
the cross sections with and without
two-neutron transfer are quite similar.
After the $2n$-transfer (before the capture)
in the reactions
$^{18}$O($\beta_2=0.1$) + $^{92}$Mo($\beta_2=0.05$)$\to ^{16}$O($\beta_2=0$) + $^{94}$Mo($\beta_2=0.151$),
$^{18}$O($\beta_2=0.1$) + $^{74}$Ge($\beta_2=0.283$)$\to ^{16}$O($\beta_2=0$) + $^{76}$Ge($\beta_2=0.262$),
$^{18}$O($\beta_2=0.1$)+$^{112}$Sn($\beta_2=0.123$)$\to ^{16}$O($\beta_2=0$)+$^{114}$Sn($\beta_2=0.121$),
$^{18}$O($\beta_2=0.1$)+$^{118}$Sn($\beta_2=0.111$)$\to ^{16}$O($\beta_2=0$)+$^{120}$Sn($\beta_2=0.104$),
and
$^{18}$O($\beta_2=0.1$)+$^{124}$Sn($\beta_2=0.095$)$\to ^{16}$O($\beta_2=0$)+$^{126}$Sn($\beta_2=0.09$)
the deformations of the nuclei  decrease  and
the values of the corresponding Coulomb barriers  increase.
As a result, the transfer
suppresses the capture process at the sub-barrier energies.
The suppression becomes  stronger with decreasing  energy.
As examples, in Fig.~1 and 2 we show this effect for the reactions
$^{18}$O+$^{74}$Ge,$^{92}$Mo.

The quantum diffusion approach was applied to study
the role of the neutron transfer with   positive $Q$-value
in the capture  reactions
$^{18}$O+$^{50}$Cr,
%$^{18}$O+$^{58}$Ni,
$^{18}$O+$^{74}$Ge,
$^{18}$O+$^{92}$Mo, and
$^{18}$O+$^{112,118,124}$Sn at sub-barrier energies.
We found that the change of the magnitude of the
 capture cross section after the neutron transfer
occurs due to the change of the deformations of  nuclei.
The  effect of the  neutron transfer is an indirect effect of the quadrupole
deformation.
If in the reaction under consideration  the deformations of nuclei
decrease after the neutron transfer, the neutron transfer   suppresses the capture cross section. As shown,
the capture cross sections for the reactions
$^{16}$O+$^{52}$Cr,$^{76}$Ge,$^{94}$Mo,$^{114,120,126}$Sn and
$^{18}$O+$^{50}$Cr,$^{74}$Ge,$^{92}$Mo,$^{112,118,124}$Sn are almost coincide, respectively.

This work was supported by DFG, NSFC, and RFBR.
The IN2P3(France) - JINR(Dubna)
%, MTA(Hungary)-JINR(Dubna)
and Polish - JINR(Dubna)
Cooperation Programmes are gratefully acknowledged.\\

\end{document}